\documentclass{ws-procs9x6}

\begin{document}

\title{ K$^{*0}$ Photoproduction off the proton\\ at CLAS} 

\author{I. Hleiqawi and K. Hicks\footnote{\uppercase{F}or
  \uppercase{CLAS} \uppercase{C}ollaboration.}}

\address{ Department of Physics and Astronomy,\\ 
OHIO UNIVERSITY,\\
Athens, OH 45701., USA\\
E-mail: hleiqawi@jlab.org}
 
\maketitle

%%%%%%%%%%%%%%%%%%%%%%%%%%%%%%%%%%%%%%%%%%%%%%%%%%%%%%%%%%%%%%
\newcommand{\p}{{\rm p}}
\newcommand{\n}{{\rm n}}
\newcommand{\e}{{\rm e}}
\newcommand{\K}{{\rm K}}
\newcommand{\Y}{{\rm Y}}
%%%%%%%%%%%%%%%%%%%%%%%%%%%%%%%%%%%%%%%%%%%%%%%%%%%%%%%%%%%%%%s
%%%%%%%%%%%%%%%%%%%%%%%%
\abstracts{
We report here differential cross sections for the reaction 
$\gamma p \to  K^{*0}\Sigma^+$ for the first time with 
high statistics. The measurements were done using 
the CEBAF Large Acceptance Spectrometer (CLAS) at 
Jefferson Lab. Data were taken at 3.115 GeV electron beam energy 
using tagged photons incident upon a liquid hydrogen target. 
In addition, our analysis provides information about nucleon 
resonances and their couplings to $K^*$ decay. 
Our data will be used to test predictions of theoretical 
models of K$^*$ production, and to provide constraints 
on the $K^*\Sigma N$ coupling constant used in these models.
}
%%%%%%%%%%%%%%%%%%%%%%%%%%%%%%%%%%%%%%%%%%%%
\section{ Introduction }\label{sec:intro}
%%%%%%%%%%%%%%%%%%%%%%%%%%%%%%%%%%%%%%%%%%%%
Some Quark Models predict yet undetected baryon 
resonances [\refcite{isgur1}]. These ``{\it missing}" resonances 
have not been observed via strong interactions and could be 
observed via the photoproduction method. Of special interest are 
the nucleon resonances N$^*$ which could couple strongly to 
$K\Lambda$ and $K\Sigma$ [\refcite{capstick2}]. Moreover, higher 
mass nucleon resonances could favor decaying into $K^*\Sigma$, 
near threshold. Vector meson electro- and photoproduction near 
threshold might provide good knowledge about these resonances, 
their internal structure, and their couplings to vector mesons. 
This has been the main motivation for studying strange mesons 
photoproduction off the proton, $\gamma\p \rightarrow \K\Y$, 
where Y denotes the hyperon,
\begin{equation}   
        \label{Eq:strangeness}             
  \gamma\p \rightarrow K^+\Lambda, ~~~~~~~~
  \gamma\p \rightarrow K^+ \Sigma^0, ~~~~~~~~   
  \gamma\p \rightarrow K^{*0}\Sigma^{+}.  
\end{equation}      
Analyses and results on the first two reactions have been 
published [\refcite{{McNabb2},{bradford}}]. However, the third 
channel has not been studied due to its small cross section 
and the difficulty of detecting kaons. The availability of the 
high intensity electron facility and the CEBAF Large Acceptance 
Spectrometer (CLAS), at Jefferson Lab, has made it possible to 
study this channel.

{\it Theoretical Model:}
%%%%%%%%%%%%%%%%%%%%%%%%%%%%%%%%%%%%%%%%%%%%
a theoretical quark model [\refcite{zhao}] has been developed 
to study nucleon resonances and to present quark model predictions 
for the K$^{*0}$ production. In addition to using common quark model 
parameters, this model uses two free parameters: the vector and 
tensor couplings, {\it a} and {\it b}, for the quark-K$^*$ interaction.
These are the basic parameters in the model and are related to the 
K$^*\Sigma$N$^*$ couplings that appear in the quark model symmetry 
limit. The SU(3)-flavor-blind assumption of non-perturbative QCD is 
adopted, which suggests the above two parameters should 
have values close to those used in the $\omega$ and $\rho$  
photoproduction. That is, the model predicts cross sections for 
the K$^{*0}$ production using couplings extracted from non-strange 
production. Our data will provide a good test of this model.

K$^*$ production is related to other strangeness production in
Eq.~(\ref{Eq:strangeness}). At the {\it hadronic level}, 
($\gamma,K^*$) and ($\gamma,K$) are related to each other since 
one reaction contains meson production in the other process as the 
$t$-channel exchanged particle, and therefore constrains the range of 
available couplings. This allows both reactions to use the same 
observables. At the {\it quark level}, both reactions involve the 
creation of s$\bar{s}$ pair from the vacuum.  Hence, these 
reactions are related also through the quark model.
%%%%%%%%%%%%%%%%%%%%%%%%%%%%%%%%%%%%%%%%%%%%
\section{ Experiment and Analysis}\label{sec:exp}
%%%%%%%%%%%%%%%%%%%%%%%%%%%%%%%%%%%%%%%%%%%%
The K$^{*0}$ photoproduction data were extracted from the ``g1c" data 
set (4.5 billion triggers) using the CLAS detector [\refcite{clas}], 
at Jefferson Lab's Hall B. Data for our analysis were taken at 3.115 
GeV electron beam energies.

The photon beam was produced in Hall B photon 
tagger [\refcite{sober}], by directing the electron beam toward 
the tagger's radiator where the beam strikes an Al foil creating 
bremsstrahlung photons. Each photon was tagged by measuring energies 
of the recoiling electrons in the tagger spectrometer, up to 95\% of
incident electron beam energy. The photon beam was then directed to 
hit the target at the heart of CLAS, through a pair of collimators 
which trim the beam halo. The target, a liquid hydrogen of length 
17.85 cm, was surrounded by a start counter which provides prompt 
timing measurements once photons interact with the target.
A coincidence between the tagger and the start counter 
provided the trigger. See Ref.~[\refcite{clas}] for details
on CLAS spectrometer.

\begin{figure}[h]
\vspace{22mm}
%\epsfxsize=20cm   %width of figure - will enlarge/reduce the figures
%\epsfbox{01feyn_dia.ps}
%\figurebox{2cm}{3cm}{} %to have a box alone 
%\centerline{\epsfxsize=5.5in\epsfbox{01feyn_dia.ps} }  
\centering{\includegraphics{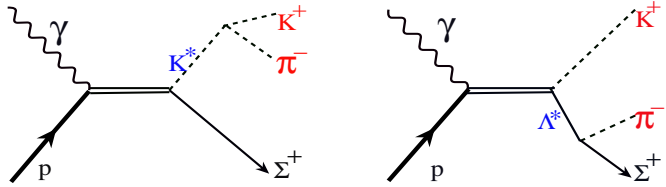}}
\caption{The reaction of interest (left) and 
	   a background channel, $\Lambda$(1520), Eq.~(\ref{Eq:bg}).       
\label{feynman}}
\end{figure}

When the $\gamma$ strikes the $p$, in addition to K$^{*0}$ other 
background channels are produced from K$^{+}$ production,
\begin{equation}
  \label{Eq:bg}
      \gamma\p  \to \K^+\Sigma^0,  ~~~~~~~~~
      \gamma\p  \to \K^+\Lambda/\Y^*  
\end{equation} 
In this experiment, K$^{*0}$ decays to K$^+\pi^-$, while 
both ground states
% \footnote{
%        $\Sigma^0 \to \Lambda\gamma$ (100\%)\\
%	$\Lambda \to \pi^-\p$ (64\%)
%			} 
$\Sigma^0$(1193) and $\Lambda$(1116), in the background channels 
Eq.~(\ref{Eq:bg}), decay (with 64\% probability) to $\pi^-\p$, and 
the $\Lambda$(1520) decays (with 14\% probability) into 
$\pi^-\Sigma^+$. In addition, we have other backgrounds, from higher 
excited states of $\Lambda$. Fig.~(\ref{feynman}) shows diagrams of 
the K$^{*0}$ (signal) and $\Lambda$(1520) production. 
%%%%%%%%%%%%%%%%%%%%%%
%%%%%%%%%%%%%%%%%%%%%%

The only detected particles are K$^+$ and $\pi^-$.
The K$^{*0}$ meson is identified from its decay products, 
K$^+\pi^-$, while $\Sigma^+$ hyperon is reconstructed from 
the missing mass of K$^{*0}$. We have two major background 
contributions: (i) from misidentified $\pi^+$  as a K$^+$, 
and (ii) from the K$^+$ production, Eq.~(\ref{Eq:bg}). 
Contributions from $\Sigma^0$(1193) and  $\Lambda$(1116) hyperons 
were removed by applying a cut on the K$^+$ missing mass, MM(K$^+$). 
To remove contributions from Y$^*$ states and the misidentified 
pions, and to identify $\Sigma^+$ as well as to extract the yields 
from it, we applied two cuts on K$^{*0}$ mass spectrum: 
a peak cut centered at 0.892 GeV (the K$^{*0}$ cut) and side-band cuts 
(the Y$^{*}$ cut) on either side of K$^{*0}$ peak. Using the peak cut, 
our signal (the $\Sigma^+$ yield) is reconstructed from the missing mass, 
MM(K$^+\pi^-$), and is fit with a gaussian plus a polynomial fit to the 
background. On the other hand, using the side-band cut, the Y$^*$ 
{\it physics} background is calculated from the MM(K$^+\pi^-$), 
obtaining another $\Sigma^+$ peak that is fit in the same way. 
From these fits we obtained two $\Sigma^+$ yields, one from K$^{*}$ 
cut and the other from Y$^{*}$ cut. By subtracting the latter 
from the former one we obtained the final $K^{*0}\Sigma^+$ yields. 
The measured yields were then normalized by the real photon flux and 
corrected for the CLAS detector acceptance (from Monte Carlo).
%%%%%%%%%%%%%%%%%%%%%%
\section{Results and conclusions}\label{sec:results}
\begin{figure}[t]
\vspace{36mm}
%\epsfxsize=6.2cm   %width of figure - will enlarge/reduce the figures
%\epsfbox{070dxsec_vs_e_stand_c.ps}
%\figurebox{2cm}{3cm}{} %to have a box alone 
\centering{\includegraphics{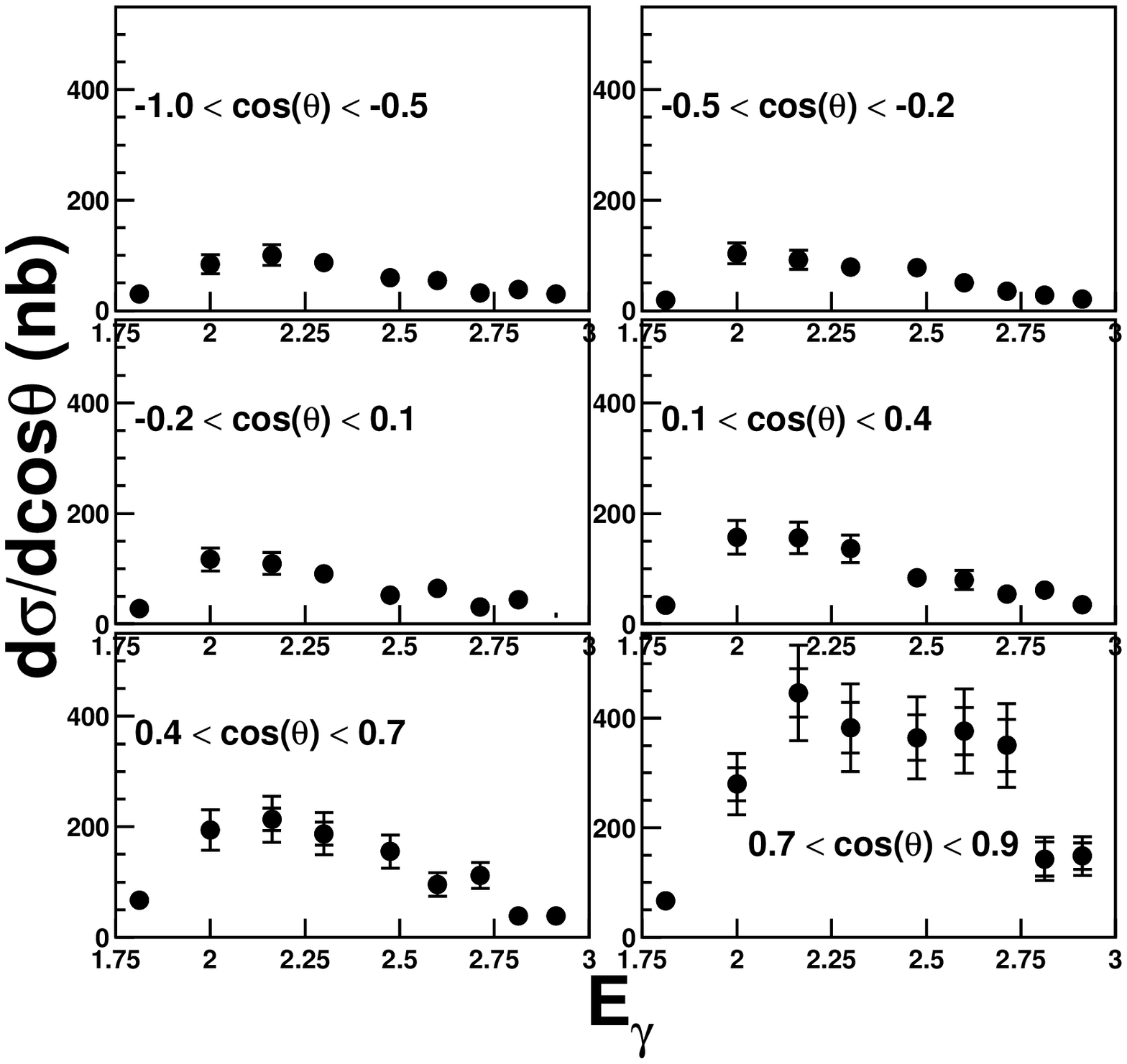}}
\centering{\includegraphics{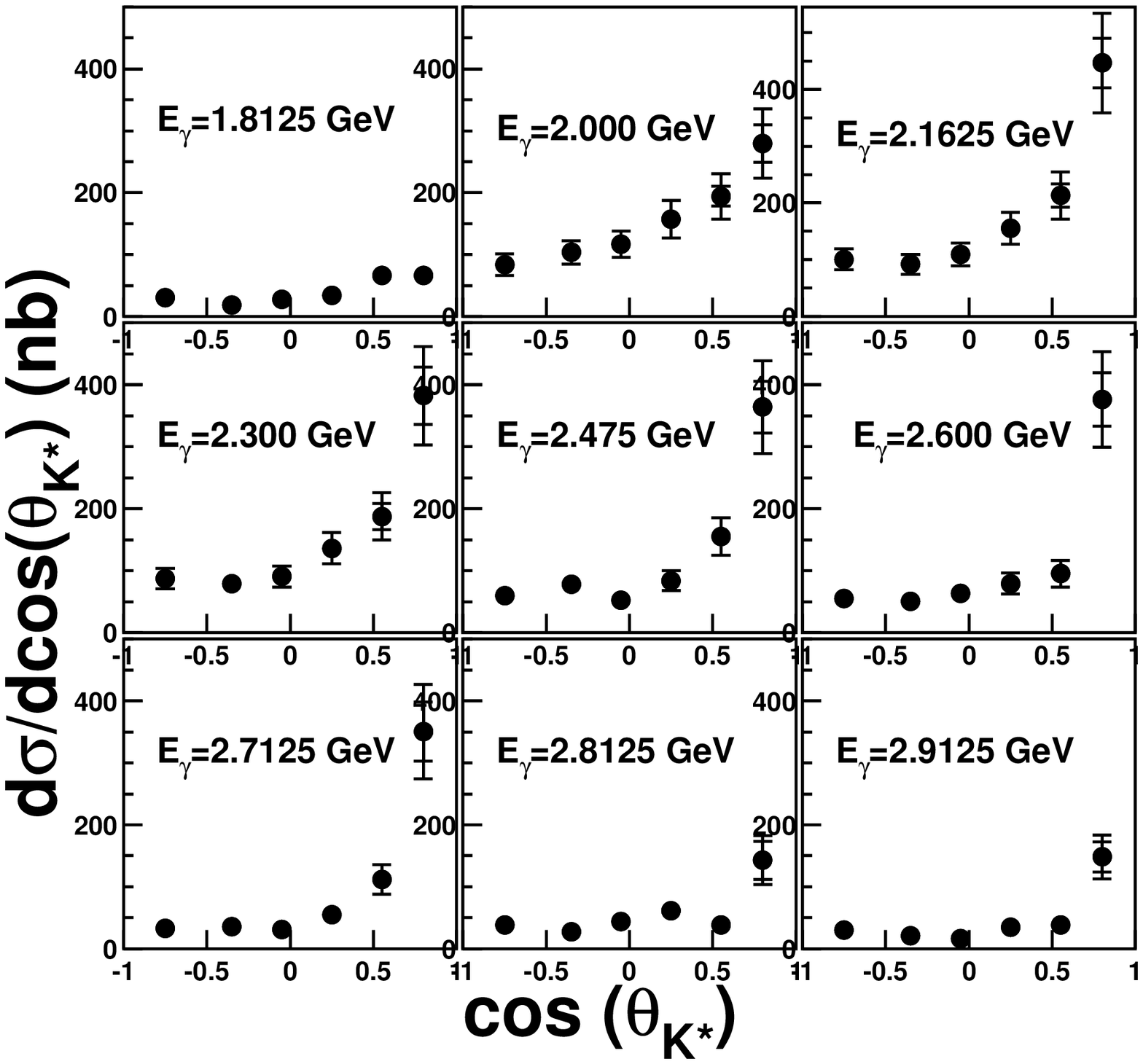}}     
\caption{Differential cross sections for 
          $\gamma$p $\rightarrow$ K$^{\rm *0}$$\Sigma^+$.
          Energy evolution for six angular bins (left) and 
	  angular distribution for nine photon energy bins (right).      
\label{xsec}}
\end{figure}

%%%%%%%%%%%%%%%%%%%%%%
Fig.~(\ref{xsec}) shows the differential cross sections for 
the K$^{*0}$ photoproduction. Although not shown in this figure, 
the cross sections are in good agreement with the theoretical 
model of Zhao [\refcite{zhao}], after a small modification of 
the K$^*$-quark vector and tensor coupling constants (see Section 1).
More details will be presented in an upcoming paper.  This suggests 
that the theoretical model of Zhao for vector meson production 
within the quark model has some predictive power.

At small angles, our data show the cross section is dominated by 
$t$-channel exchange except at the highest photon energy bin. 
At higher energies, the forward-peaking moves out of our 
detector acceptance, as CLAS can only measure the  
production angle up to cos($\Theta_{K^{*0}}^{c.m.}$) $ < $ 0.9 
due to the beam-pipe hole through the center of CLAS. 
%Total cross sections will be estimated using simulations 
%that include $t$-channel exchange to fill this forward-angle gap.

%%%%%%%%%%%%%%%%%%%%%%

\section*{Acknowledgments}
We would like to thank A. Tkabladze for his help in this project, and also 
to C. Bennhold and Q. Zhao for their help and for valuable discussions.

\end{document}